

\input phyzzx

\def\ls#1{_{\lower1.5pt\hbox{$\scriptstyle #1$}}}

\frontpagetrue


\let\picnaturalsize=N
\def\picsize{1.0in}
\def\picfilename{scipp_tree.eps}

\let\nopictures=Y

\ifx\nopictures Y\else{\ifx\epsfloaded Y\else\input epsf \fi
\let\epsfloaded=Y
{\line{\hbox{\ifx\picnaturalsize N\epsfxsize \picsize\fi
{\epsfbox{\picfilename}}}\hfill\vbox{


\hbox{SCIPP 94/01}
\hbox{January 1994}
\vskip1.2in
}
}}}\fi


\def\SCIPP{\centerline {\it Santa Cruz Institute for Particle Physics}
  \centerline{\it University of California, Santa Cruz, CA 95064}}
\overfullrule 0pt

\pubtype{ T}     

\hfill\hbox{SCIPP 94/01}

\hfill\hbox{January 1994}

\vskip1.in

\title{{Electroweak Baryogenesis in the Adiabatic Limit}
\foot{Supported in part by the U.S. Department of Energy
and the Texas National Research Laboratory Commission under grant
numbers RGFY 93-263 and RGFY 93-330.}}
\author{Michael Dine and Scott Thomas
}
\SCIPP
\vskip1cm
\vbox{
\centerline{\bf Abstract}

\parskip 0pt
\parindent 25pt
\overfullrule=0pt
\baselineskip=18pt
\tolerance 3500
\endpage
\pagenumber=1
\singlespace
\bigskip

Electroweak baryogenesis can occur in the ``adiabatic limit,''
in which expanding bubbles of true vacuum are assumed to have
rather thick walls and be slowly moving.  Here the problem of
calculating the baryon asymmetry in this limit is reconsidered.
A simple  prescription for obtaining the relevant kinetic equations
is given. An additional suppression beyond that usually assumed is found.
This arises because the generation of an asymmetry requires
violation of approximately conserved currents by Higgs expectation
values, which are small near the front of the walls.
As applications, the asymmetries in multi-Higgs models and the
minimal supersymmetric standard model are estimated to be
proportional to $\alpha_w^8$ and $\alpha_w^6$ respectively.
Also, in this limit the baryon asymmetry in the Minimal Standard Model
is extraordinarily small.
}

\vfill
\submit{Physics Letters B}
\vfill

\singlespace
\bigskip

\baselineskip=20pt

\chapter{Introduction}

\REF\sakharov{A. Sakharov, JETP Lett. {\bf 5} (1967) 24.}

\REF\krs{V. Kuzmin, V. Rubakov, and M. Shaposhnikov,
Phys. Lett. B {\bf 155} (1985) 36.}

\REF\cknreview{For an excellent review, see
A. Cohen, D. Kaplan, and A. Nelson,
UCSD preprint UCSD-PTH-93-02 (1993), to
appear in Annual Review of Nuclear and Particle Science,
{\bf 43} (1993).}

\REF\farrar{G. Farrar and M. Shaposhnikov,
CERN preprint CERN-TH-6734/93 (1993).}

The origin of the baryon asymmetry
of the universe is a long-standing problem.\refmark{\sakharov}
With the recognition that baryon number is badly violated
in the standard model
at high temperatures,\refmark{\krs} there have been
many proposals for how an asymmetry might be generated
at the electroweak phase transition.\refmark{\cknreview}
Virtually all of these proposals assume that the
electroweak phase transition is first order, providing the
required departure from equilibrium.  CP violation
already exists in the minimal standard model (MSM);
extensions of the standard model
generically contain further
sources of CP violation.
Since all three generations of quarks
must be involved, any asymmetry which arises
in the MSM should be extremely small.
Most studies of electroweak baryogenesis have
involved extensions of the MSM.
However, there
have been a number of attempts to
explain the asymmetry within
the framework of the MSM, the most recent being
that of ref. \farrar.

\REF\mvst{L. McLerran, M. Shaposhnikov, N. Turok
and, M. Voloshin, Phys. Lett. B {\bf 256} (1991) 451.}

\REF\dhss{M. Dine, P. Huet, R. Singleton, and L. Susskind,
Phys. Lett. B {\bf 256} (1991) 351.}

\REF\cknsg{A. Cohen, D. Kaplan and A. Nelson, Phys. Lett. B
{\bf 263} (1991) 86.}

\REF\cknthinwall{A. Cohen, D. Kaplan and A. Nelson, Nucl. Phys. B
{\bf 373} (1992) 453.}

\REF\turoka{N. Turok and J. Zadrozny, Phys. Rev. Lett. {\bf 65}
(1990) 2331.}

\REF\turokb{D. Grigoriev, M. Shaposhnikov, and N. Turok, Phys.
Lett. B {\bf 275} (1992) 395.}

The electroweak transition, if first order, proceeds by formation
of bubbles.
The baryon asymmetry is produced in or
near the bubble walls, which are the sites
of the most significant departures from equilibrium.
The various schemes considered
to date for producing an asymmetry divide into
two classes.
The first is referred to
as the ``adiabatic case."
If a wall is slowly moving and thick, its passage
through the plasma is nearly adiabatic, in the sense that
all quantities (but the baryon number) should be close to equilibrium.
In this case, the time-varying Higgs field can act to bias the
baryon-violating processes.\refmark{\mvst,\dhss,\cknsg}
In the second scheme, the wall is assumed to be thin
and rapidly moving, so more significant departures
from equilibrium can occur.  Cohen, Kaplan and Nelson
have pointed out that in this limit, scattering of particles from
the wall
can produce an asymmetry in front of
the wall for some (approximately conserved) quantum
number.  This, in turn, can bias the baryon violating
processes.\refmark{\cknthinwall}
Other non-adiabatic
scenarios have been considered in refs. \turoka\ and \turokb.

\REF\sg{A. Cohen and D. Kaplan, Phys. Lett. B
{\bf 199} (1987) 251; Nucl. Phys. B {\bf 308} (1988) 913.}

In this paper, certain aspects of the
adiabatic case will be considered.
The
adiabatic picture
has been most carefully developed in refs. \dhss\ and
\cknsg; it is closely
related to the ``spontaneous baryogenesis" scheme
of Cohen and Kaplan.\refmark{\sg}
In ref. \dhss,
the problem was analyzed by obtaining the leading
coupling between the time-varying Higgs field and
Chern-Simons number.  In order to make the analysis
as simple as possible, models were considered in which this
coupling arose from loops of heavy fields (massive compared to the
temperature of the transition).   However,
while the analysis is simple in this limit, the resulting
asymmetry is quite small, for two reasons.  First, there
is simply the suppression by the masses of heavy particles.
But in addition, there is a suppression by four powers of
coupling constant, $g$.
This latter suppression results because
the operator which gives rise to the coupling is quadratic in
$g \phi$ (where $g$ denotes a gauge or Yukawa coupling and
$\phi$ denotes the Higgs field).
The
baryon number violating process cuts off at a
value of the scalar field, $\phi_{co}$ (in the
notation of ref. \cknreview), for which
$$
g \phi_{co} \sim \alpha_w T,
\eqn\phicutoff
$$
i.e. for a rather small value of the Higgs field.
There has been some debate in the literature as to how
large, numerically, this suppression may be.
This question will be taken up
later;
here the parametric dependence on the couplings, $g$, will
be determined.

In many models the coupling of the scalar field
to the Chern-Simons number arises due to light fields.
While this case is potentially
more promising, it is inherently
more complicated, involving all the subtleties of
real-time, finite temperature field theory.
In order
to deal with this, an attractive
method has been proposed by Cohen, Kaplan and Nelson in
ref. \cknsg.  The results for the asymmetry are distinctly
more promising; in particular, it is not obvious
from the analysis of ref. \cknsg~that the
suppression by powers of coupling constant
mentioned above is obtained.

However, this method has limitations.
In the multi-Higgs model treated by many authors, for example,
it cannot be valid for very small quark mass, since
a non-zero asymmetry results
even as the top quark mass
tends to zero.  This fact has already been remarked by the
authors of ref. \cknsg, who argue that for small
mass there will be suppression
by powers of Yukawa couplings.
But a non-zero asymmetry is also obtained for vanishing Higgs
expectation value,
implying an additional suppression proportional to powers
of $\phi_{co}$ has been neglected.
The method also
does not lend itself to the treatment of a number of
other interesting situations, such as baryogenesis
in the MSM.

In this note, baryogenesis in this
adiabatic, light field, case
will be reexamined.
In this discussion, $\phi_{co}$ will be assumed small.
While this leads inevitably
to rather small asymmetries, it significantly simplifies the analysis.
This is because the $B$-violating interactions switch
off rapidly, in a time of order
$$
t_{co} \sim {\phi_{co} \ell \over v \Delta \phi}\eqn\tco
$$
where $\ell$ is the wall thickness and $v$ its velocity, and
$\Delta \phi$ is the total change in $\phi$ across the wall (typically
of order $T$).
Even if the wall is rather thick,
$t_{co} \sim T^{-1}$.
During this time, the
particle number densities can change only
by small amounts.
The kinetic equations
can then be written down and solved almost trivially.

\REF\toappear{M. Dine and S. Thomas, SCIPP preprint to appear.}

The case where the cutoff is not so small
can be analyzed by a more complete set of kinetic equations then will be
considered below.  This will be postponed
for a future publication.\refmark{\toappear}

Given the assumption of small $\phi_{co}$,
we first note that a more direct, naive
treatment of the problem leads to the sort of suppression by
powers of $\phi_{co}$ observed in the large mass case.
This naive treatment involves two steps.
The coupling of the time-varying field to the
Chern-Simons number is first computed.
A rate equation for the baryon number
is then developed
by determining
how this coupling biases the baryon-violating
rate.
We next consider the approach of ref. \cknsg, in which
the theory is rewritten so as to exhibit
the coupling of the Higgs field to certain currents.  Here the issue
will be to derive suitable rate equations.
In fact, the results of the naive treatment are recovered.

\REF\shaposhmc{J. Ambjorn, T. Askaard, H. Porter, and
M. Shaposhnikov, Phys. Lett. B {\bf 244} (1990) 479;
Nucl. Phys. B {\bf 353} (1991) 346.}

\REF\lmt{B. Liu, L. McLerran, and N. Turok, Phys. Rev. D
{\bf 46} (1992) 2228.}

\REF\vancouver{M. Dine, ``Electroweak Baryogenesis,'' in Proceedings of
the Vancouver Meeting,
edited by D. Axen, D. Bryman, and M. Comyn (World Scientific,
Singapore, 1992) p. 831.}

In the end, then, the asymmetry in the adiabatic limit
for the multi-Higgs model
is not of order
$\alpha_w^4$, but rather $\alpha_w^8$ (times $CP$-violating
phases and dynamical factors involving sphaleron
rates, the bubble
profile and velocity, etc).\foot{Here $\alpha_w$ denotes a generic
coupling; in the multi-Higgs model, for example,
this would be multiplied
by $(\lambda_t/g_w)^2$.}
While this sounds alarmingly small,
it may yet correspond to an acceptable asymmetry.
For example, general arguments give that the high temperature
baryon-violating rate goes as $\kappa
\alpha_w^4$, but inclusion (or not)
of factors of $4\pi$ is currently a matter of prejudice.
Indeed, the only calculation of this rate is that of
ref. \shaposhmc; the authors of this paper have recently
expressed skepticism as to the validity of their
result.\refmark{\farrar}  As for the significance of the
suppression, some authors have argued,
by interpolating a formula valid for large $\phi$, that\refmark{\lmt}
$$
g \phi_{co} \sim 7 \alpha_w \sim 0.2~T.
\eqn\cutoff
$$
One of us has argued that formulas which interpolate between
the small and large $\phi$ limits are likely to give
$2-3$ instead of $7$ in eq. \cutoff.\refmark{\vancouver}
The same estimate, however, tends to give quite a large value
of $\kappa$.
At best, these estimates are just educated guesses.
More extensive simulations are essential to give
results reliable even as to order of magnitude.

Apart from revising earlier estimates in certain models,
the methods developed here may be applied to the MSM.
Recently it has been suggested that the MSM may lead to
a suitable baryon asymmetry.\refmark{\farrar}
Perturbative treatments of the phase transition
suggest that the wall is indeed thick and slowly moving,
so that the adiabatic treatment should not be unreasonable.
In this
case, the asymmetry turns out to be extremely small
($n_B/s \sim 10^{-32}$ or so).  In ref. \farrar, however, it
was argued that perturbation theory is not a reliable guide to
the physics of the phase transition, and that the wall is fast
and thin.  We will comment on this possibility, but are not
in a position to make any definitive statements.

%
%

\chapter{Naive Treatment of the Multi-Higgs Model}

\FIG\thetacoupling{Leading diagram which couples the phase, $\theta_1$,
to the Chern-Simons number.}

\REF\finitet{See for example S. Abel, W. Cottingham,
and I. Whittingham, Nucl. Phys. B {\bf 410} (1993) 173.}

For definiteness, following ref. \cknsg, a multi-Higgs
model with a CP violating Higgs potential will be considered.
Only one Higgs, $\phi_1$, couples to quarks
(to avoid flavor-changing neutral currents).
In the bubble wall, $\phi_1 = \rho_1(x) e^{i \theta_1(x)}$,
where $\theta_1(x)$ arises from the CP violation.
The coupling of $\theta_1$ to $F \tilde F$
may be calculated
from the
triangle diagram of fig. \thetacoupling.
The propagators in the loop are understood
as suitable finite-temperature Green's functions. The
real-time expression can be
obtained by analytic continuation of the Euclidean time
result.  In the present case, this is straightforward.
The diagram is perfectly finite in the infrared and ultraviolet,
and for slowly varying $\theta_1$ (compared
to $T^{-1}$) can be expanded in a power series in the momenta.
The result is necessarily
quadratic in $m_t = \lambda_t \rho_1$
because of the required chirality flip
($m_t$ is the effective $t$
quark mass, $\lambda_t$ is the $t$ quark Yukawa coupling).
The calculation is quite straightforward (related
calculations, for example, have been performed in ref. \finitet) and
one
obtains\foot{Ref. \finitet\ indeed calculates many of the couplings
necessary for the analysis.  However, they use the results
of ref. \cknsg\ without modification
to obtain rate equations, and thus their final results
differ from ours.}
$$
{\cal L}_{\theta}= a\theta_1{ m_t^2 \over  T^2}
{F \tilde F \over 32 \pi^2}+
{\cal O} \left( {m_t\over T} \right)^4
\eqn\thetaffcoupling
$$
where $a={14 \over 3 \pi^2} \zeta(3)$.

How might this coupling bias the sphaleron process?
Integrating by parts, ${\cal L}_{\theta}$
may be written in terms
of the Chern-Simons number
$$
{\cal L}_{\theta}= a \left( {m_t^2 \over T^2} \right)
\partial_o \theta_1 n_{cs}.
\eqn\ncstheta
$$
Suppose that the coefficient of $n_{cs}$ here is very slowly changing
with time.
This gives, effectively, a chemical potential for
Chern-Simons number,
$$
\mu_{cs}= a \left( m_t^2/T^2 \right)
 \partial_o\theta_1.
\eqn\cschemical
$$
In an (anti-)sphaleron transition, $n_{cs}$
changes by (-1)1.  Suppose $\Gamma$ is the rate for transitions changing
$n_{cs}$ at equilibrium.   Considerations of detailed balance then give that,
when the particle densities are small compared with the equilibrium
values
(as is the case for our assumption of short times),
the difference in rates for transitions changing $n_{cs}$ by
$+1$ and $-1$ is
$${dn_{cs} \over dt}= \mu_{cs}\beta \Gamma\eqn\rateeqn$$
The corresponding change in baryon number is three times larger.

This result will be obvious to many readers, but a brief
description of the derivation is perhaps useful.
Near equilibrium, for small number densities,
states differing in $n_{cs}$ by one unit differ in free energy
by an amount $\mu_{cs}$.  Yet at equilibrium, the rate of
transitions increasing and decreasing the free energy
must be equal.  Thus the ratio of these rates must equal
the ratio of Boltzmann factors for the two states,
$$e^{-\beta \mu}.\eqn\freeratio$$
For small $\mu$, this just gives eq.. \rateeqn.

Since $\mu \propto m_t^2 \propto \rho_1^2$,
this
treatment should give a baryon number proportional to $\phi_{co}^2$,
i.e. the small sort of rate discussed earlier.
To make an estimate, we make the simplifying assumption of
$\Gamma = \Gamma(\phi)$.  We then make
the further simplification of taking
$\Gamma=\kappa (\alpha_w T)^4$ for $\phi < \phi_{co}$,
and $\Gamma=0$ for $\phi > \phi_{co}$.  This
gives
$$
n_B \sim 3
a\kappa \alpha_w^4 \lambda_t^2
\phi_{co}^2 T
\Delta \theta_1.
\eqn\firstestimate
$$
Here, $\Delta \theta_1$ is the value of the $CP$-violating
phase when the baryon-violating process turns off.
The baryon to entropy ratio is of order
$$
{n_B \over s} \approx
{\kappa} \left( {100 \over g^*} \right)
\left( {\Delta \theta_1 \over \pi} \right)
\left( {\lambda_t \phi_{co} \over T} \right)^2
\times (1 \times 10^{-7}).
\eqn\roughestimate
$$
It should be noted that in the multi-Higgs case, $\Delta \theta_1$,
is itself of
order $(\phi_{co} / T )^2$ (times coupling constants), since
in the absence of quartic couplings
the
Higgs potential is $CP$-conserving.

The worst-case scenario here, in which $\kappa \sim 1$ and
$g\phi_{co} \sim \alpha_w T$, gives an unacceptably small
result, of order  $10^{-13}$ for
$n_B / s$.
As remarked above,
however, it is conceivable that $\kappa$ is large,
and $\phi_{co}$ is $3$ to $7$ times as large, so the
final result could well be large enough, provided CP-violating
phases are large.

Finally, we can ask how other processes
affect the final asymmetry.  For example, processes involving scattering
of top quarks and Higgs fields, and QCD sphaleron processes,
will change the numbers of left and right-handed fields.
Each fermion species obeys a rate equation of the form
$$
{dn_i \over dt} \simeq \beta \mu_{cs}d_i \Gamma + \gamma_i
\eqn\indivrates
$$
Here the first term represents the sphaleron process;
$d_i=0$ for $SU(2)$ singlets, $d_i=1/2$ for doublets.
As before, $\Gamma$ is the sphaleron rate, while $\gamma_i$ denotes
other processes which change the
number densities.
If $\gamma_i$
can be neglected,
then summing over the individual rates weighted with the baryon number,
eq. \indivrates\ reproduces eq. \rateeqn.

\REF\guidice{G. Giudice and M. Shaposhnikov,
CERN preprint CERN-TH.7080/93 (1993).}

The estimate described
above will be valid provided that $\gamma_i$ and $n_i$
are not too large.
In order to understand this criterion, consider a particular
process which has been discussed recently:  strong sphalerons in the
two Higgs model.\refmark{\guidice}
In this case,
the associated terms in the rate equations
are calculated just as for
the weak sphalerons.  The triangle diagram with the
$W$ bosons replaced by gluons gives rise to a chemical potential
for ``strong Chern-Simons number'', $n_{cs}^{c}$ precisely as
for $n_{cs}$.  In
this case, one obtains in the effective lagrangian
$$
{\cal L}^c_{\theta}= a^c\theta_1{ m_t^2 \over  T^2}
{G \tilde G \over 32 \pi^2}+
{\cal O} \left( {m_t\over T} \right)^4
\eqn\thetaffcouplingqcd
$$
where $a^c=-{14 \over  \pi^2} \zeta(3)$.  In other words,
a chemical potential for the QCD Chern-Simons
number results which is three times as large as that for
$SU(2)$.   The rate equation for small densities is now
$$
{dn_i \over dt} \simeq \mu_{cs}\beta
 (d_i \Gamma -3
{}~h_i \gamma)
\eqn\strongsphaleron $$
where
$h_i=1/3$ for color triplets and anti-triplets, $h_i=0$ for leptons,
and
the rate, $\gamma = \kappa^{c}(\alpha_s T)^4$.
In this equation, terms on the right
hand side linear in the densities have been neglected;
this is
correct provided none of the densities are of
order $\mu_{cs} T^2$.
This equation can be integrated, as before,
up to times for which $\phi=\phi_{co}$.
As argued above
this corresponds to a time, $t_{co} \sim T^{-1}$
(eq. \tco).
Apparently
the strong sphaleron term can be neglected
unless $\kappa^{c}$
is extremely large, of order $10^5$ or so (the precise value depending
also on $\phi_{co}\over \Delta \phi$.)
The earlier
estimate of the baryon number is then unmodified.
Other processes may be treated
similarly, such as top quark
scattering from Higgs bosons.  Again, provided the estimate
for $t_{co}$ is reasonable, there is no effect.
The case
with more complete rate equations
will be discussed
in a subsequent publication.\refmark{\toappear}


\chapter{Spontaneous Baryogenesis}

Ref. \cknsg\ suggested a different treatment, which avoids
considering directly the coupling to Chern-Simons
number.  These authors perform an anomaly-free
redefinition of the fermion fields
which eliminates
the phase from the fermion mass terms.  For example, consider
again the multi-Higgs model,
in the version where only one Higgs field couples
to ordinary quarks and leptons ($\phi_1$).
Transforming
each fermion by a phase proportional to its hypercharge
(``fermionic hypercharge", $\tilde Y$)  eliminates the
phase, $\theta_1$, from the Yukawa couplings
at the price of a coupling
$$
\partial_{\mu} \theta_1 j_{\tilde Y}^{\mu}.
\eqn\inducedcoupling
$$
This appears to have induced a chemical potential for
fermionic hypercharge,
and to bias the sphaleron process.
Suppose the fermionic part of the free energy is minimized subject
to the constraints of charge conservation and separate
$B-L$ conservation, and including this chemical potential.
The minimum lies at a nonzero value of the baryon number.
Detailed balance arguments similar to those given above
yield an equation for the rate of change of baryon number.
Even without writing this equation however, it is clearly
not the one above.
In particular,
it does not involve the modulus of the Higgs field,
$\rho_1$, at all.


To understand this question better, focus again on
short times.  In this limit, as before,
the non-zero densities of various species
may be neglected in writing
kinetic equations.
But in this limit, it is clear that if
the (small) Higgs vev is ignored, a
chemical potential for hypercharge cannot bias the
sphaleron process, since the sphaleron process does not
violate hypercharge.
Moreover, scattering of top
quarks from the Higgs particles in the plasma
(as suggested in refs. \cknreview~and \cknsg)
cannot
help, even if the scattering is rapid.
The problem is that the field redefinition by
fermionic hypercharge, while removing phases from the
fermion mass terms, induces phases in the Yukawa couplings
of the fermions to the fluctuating part of the Higgs field.
To avoid these, write the Higgs field as
$$
\phi_1 = (\rho_1 + \phi_1^{\prime})e^{i
\theta_1}
\eqn\newphiparam
$$
where $\phi_1^{\prime}$ represents the (complex) fluctuating field.
Neglecting $\rho_1$, the lagrangian in terms of these fields
contains the phase, $\theta_1$, only in the coupling
$$\partial_{\mu} \theta_1 j_Y^{\mu}$$
where $j_Y^{\mu}$ represents the full hypercharge current,
including the scalar
parts.  But this current is conserved in any process
(in the limit of small $\rho$);
the chemical
potential has no effect.

In order to obtain any asymmetry at small times terms involving
$\rho_1$ must be included.
In this case, the results of
the
naive analysis are recovered.
In particular, it is no longer true that $j_Y^{\mu}$ is
conserved; instead, at tree level
$$\partial_{\mu}j_Y^{\mu}= m_t \bar t i \gamma_5 t + \dots \eqn\jydiv$$
In order to understand how much this is violated in
the presence of background
gauge fields consider, again, a finite-temperature
Feynman diagram.  The calculation is identical to that encountered earlier,
and gives
$$\partial_{\mu}j_Y^{\mu}= {a m_t^2 \over T^2} {F \tilde F
\over 32 \pi^2}\eqn\modifieddiv$$
We interpret this as meaning, on average, the violation of hypercharge in
a sphaleron transition is
$${a m_t^2 \over T^2} \Delta n_{cs}.\eqn\avgviolation$$
So the change in free energy, on average, in a sphaleron transition   is
precisely that encountered in the naive treatment.
The rate equation obtained is thus identical.
Although the discussion give here is for the particular hypercharge
field redefinition of ref. \cknsg, the results are more
generally applicable to any model.\refmark{\toappear}

\chapter{Spontaneous Baryogenesis in the MSM and the MSSM}

\REF\phasetransition{M. Dine, R. Leigh, P.
Huet, A. Linde, and D. Linde, Phys. Rev. D {\bf 46} (1992) 550.}

\REF\mt{B. Liu, L. McLerran, and N. Turok,
Phys. Rev. D {\bf 46} (1992) 2268.}

\REF\bubblewall{P. Huet, R. Leigh, K. Kajanti, B. Liu, and
L. McLerran, Phys. Rev. D {\bf 48} (1993) 2477. }

\REF\gaillard{J. Ellis and M. Gaillard, Nucl. Phys. B
{\bf 150} (1979) 141; I. Khriplovich and A. Vainshtein,
Minn. preprint TPI-MINN-93/91 (1993).}


A naive, perturbative treatment of the electroweak phase
transition in the MSM, for Higgs masses larger $25$ GeV
or so (well below the current LEP limits, of course), gives
a bubble wall which is thick and rather slowly moving,
and therefore in the adiabatic
regime.\refmark{\phasetransition-\bubblewall}
Having acquired some confidence in our understanding of
spontaneous electroweak baryogenesis, the case
of the minimal standard model may be considered.
In the spirit of the naive treatment,
the coupling of the Higgs field to Chern-Simons number
must be found.
Assuming $g \phi_{co} \sim \alpha T$,
the leading operator should be one
with a minimal number
of external Higgs fields (each additional loop costs
roughly a factor of $\alpha$, whereas a pair of external
Higgs fields costs  a factor of $\alpha^2$).  The simplest
such operator is
$$
{\cal L}_{\rm MSM}= {\gamma \vert \phi \vert^2 \over T^2} F \tilde F.
\eqn\msmcoupling
$$
Such a coupling is, of course, $CP$-violating, and so
arise only at high orders.
Indeed, in a manner similar to the calculation
of $\theta$ renormalization in the standard
model,\refmark{\gaillard}
such a coupling cannot occur before 7-loop order.
Six loops
are required to obtain the Jarlskog invariant.
A seventh
loop involving a hypercharge gauge boson is also needed.
These diagrams are perfectly finite in the infrared and
ultraviolet, and
for momenta small compared to $T$,
represent the first term in a Taylor expansion of the
amplitude in powers of the momentum.
Thus no difficulty with the analytic continuation
of the result is expected, and we estimate
$$
{n_B \over s} \sim \kappa \alpha_w^6
J / g^*
\eqn\msmasymmetry
$$
where
$$J = {\rm Im}~{\rm Det} \left( \lambda_U \lambda_U^{\dagger}
{}~,~  \lambda_D \lambda_D^{\dagger}  \right)
{}~\sim 10^{-21}
$$
and $\lambda_U$ and $\lambda_D$ are the up and down type quark
Yukawa matrices.
We have not attempted to include further
suppression factors of $\pi$, etc., since
the answer is already extremely small (one might guess that
these will be at least $(2 \pi)^{-7} \sim 10^{-5}$,
since this is
effectively a $21$-dimensional Feynman integration).
So even if $\kappa$ is quite large, and the suppression described
earlier is small, the asymmetry is extremely tiny; $10^{-32}$
is probably a quite conservative estimate.

\REF\shaposhnikovew{The philosophy behind this
view is spelled out most forcefully
in M. Shaposhnikov, CERN preprint
CERN-TH.6918/93 (1993); the following paper
is argued to provide some numerical support for this
view:  K. Kajantie, K. Rummukainen, and M. Shaposhnikov,
CERN preprint CERN-TH.6901/93, 1993.}

\REF\twoloop{P. Arnold and E. Espinosa, Phys. Rev. D {\bf 47}
(1993) 3546; J. Bagnasco and M. Dine,
Phys. Lett. B {\bf 303} (1993) 308.}

\REF\gavela{M. Gavela, P. Hernandez, J. Orloff, and O. Pene,
HUTP-93/A036 (1993).}

Shaposhnikov
and collaborators have recently argued that the phase transition
is much more strongly first order than perturbation theory
suggests.\refmark{\shaposhnikovew}
This is an important ingredient in the analysis of
ref. \farrar, where it is assumed that the appropriate limit
is the ``thin-wall," highly non-adiabatic situation, in which
scattering of particles from the wall is the most important
process.  We are rather skeptical of this claim.  Moreover, a crude
estimate of the mean free path for scattering of top quarks
passing through the wall, gives a result of order $T^{-1}$,
so that even for an extremely thin wall, the scattering treatment
may not be appropriate.  However, it is a crucial assumption in
the work
of ref. \farrar\ that perturbation theory is an
extremely poor guide for all
significant questions about the
phase transition.  This may be the case (though recent
studies of two loop thermal effects have yielded
only modest corrections\refmark{\twoloop}).
Still, it would be quite amazing if
these non-perturbative effects change the final asymmetry
by $22$ orders of magnitude!  (For a critique of the calculation
of ref. \farrar\ see ref. \gavela.)

\REF\susydn{These issues have
already been encountered in ref.
\dhss\ and A. Cohen and A. Nelson,
Phys. Lett. B {\bf 297} (1992) 111.  The latter authors
were aware of the possibility of further suppressions.}

\FIG\mssmdiagram{Leading diagram in the MSSM which
couples the Higgs fields to the Chern-Simons number.}

\REF\cpatom{For the current bounds on supersymmetric phases
see for example W. Fischler, S. Paban, and S. Thomas,
Phys. Lett. B {\bf 289} (1992) 373, and refs. there in.}

We close this section by extending the earlier
estimates to the minimal
supersymmetric standard model.
The scale of superparticle masses is assumed to be
of order $T$.
The principle sources of CP violation
are assumed to lie in phases of the $\mu$ term and
soft breaking mass terms of squarks and gauginos.
The estimate is very similar to that of
ref. \dhss.
Diagrams as in fig. \mssmdiagram\
give rise to couplings such as
$$
{\cal L}_{cs} = a g^2 \sin \delta~
{H_1 H_2 \over T^2}
{F \tilde F \over 32 \pi^2}
\eqn\mssmcoupling
$$
where $\delta$ represents some
combination of CP-violating phases.
Repeating the earlier estimates, we obtain for the asymmetry
$$
{n_B \over s} \sim {\kappa} \sin \delta
\left( {100 \over g^*} \right)
\left( {g \phi_{co} \over T} \right)^2
\times (8 \times 10^{-8})
\eqn\mssmroughestimate
$$
Here there is an additional difficulty.
{}From limits on
electric dipole moments (edms) of atoms and the neutron
it is expected that
$\sin \delta <10^{-2} - 10^{-3}$.\refmark{\susydn,\cpatom}
Again one must be fortunate
with $\kappa$ and $\phi_{co}$ in order to obtain an acceptable
asymmetry.  Still, without further simulations, it is hard to
rule out this possibility.  A supersymmetry aficionado might
even view this estimate as tantalizingly close, and suggestive
that edms should not be far below the
current bounds.\refmark{\susydn,\cpatom}

\chapter{Conclusions}

{}From all this, we conclude that electroweak baryogenesis,
in the adiabatic limit, is less efficient than has been widely
believed.
Due to the existence of approximately conserved quantities,
violated by Higgs expectation values,
the asymmetries which arise in the adiabatic
limit involve additional powers of couplings beyond
those usually assumed.  On the other hand, while
formal arguments can be given to
determine
the parametric dependence on
couplings, there are enormous uncertainties in the
final numerical results.  Even obtaining order of magnitude
estimates requires knowledge of the sphaleron transition rates
for small or vanishing Higgs field.  These are, by definition, not
accessible to semiclassical treatment.  Crude estimates
give widely varying answers, and existing calculations
are, according to their own authors, unreliable even
as to order of magnitude.  Clearly, improved simulations are
necessary if we are to know whether acceptable asymmetries
can be obtained in the adiabatic limit.

In the non-adiabatic, thin wall, regime, there is good
reason to believe that acceptable asymmetries can be
obtained.\refmark{\cknthinwall}  For these, $\kappa$
does not need to be large, and $\phi_{co}$ can be small,
since the asymmetry is produced in front of the wall
where the Higgs field essentially vanishes (indeed, this
issue was stressed in ref. \cknthinwall).
The MSM and MSSM are probably far from this regime,
but multi-Higgs models may well yield sufficiently
violent transitions.  If the baryon asymmetry were
produced at the electroweak phase transition,
and if the ``worst case'' scenario of small $\kappa$
and $\phi_{co}$ is correct,
a significant step beyond currently popular
theoretical ideas is likely required.

We would like to thank A. Cohen and D. Kaplan for comments on
an early version of this manuscript.

\refout
\figout
\end
\bye